\documentclass[preprint]{aastex}

\shorttitle{MIR-EXCESS OF RED EARLY-TYPE GALAXIES AT $z$ $<$ 1}
\shortauthors{Ko et al.}

\usepackage{natbib}

\begin{document}

\title{TRACING RECENT STAR FORMATION OF RED EARLY-TYPE GALAXIES OUT TO $z$ $\sim$ 1}

\author{Jongwan Ko\altaffilmark{1}, Ho Seong Hwang\altaffilmark{2}, Myungshin Im\altaffilmark{3}, Damien Le Borgne\altaffilmark{4,5}, Jong Chul Lee\altaffilmark{1},
and David Elbaz\altaffilmark{6}}

\altaffiltext{1}{Korea Astronomy and Space Science Institute, Daejeon 305-348, Republic of Korea}
\altaffiltext{2}{Smithsonian Astrophysical Observatory, 60 Garden Street, Cambridge, MA 02138, USA}
\altaffiltext{3}{Center for the Exploration of the Origin of the Universe, Astronomy Program, Department of Physics and Astronomy, Seoul National University, Seoul, Republic of Korea}
\altaffiltext{4}{Sorbonne Universit\'{e}s, UPMC Univ Paris 06, UMR 7095, Institut d'Astrophysique de Paris, F-75005, Paris, France}
\altaffiltext{5}{CNRS, UMR 7095, Institut d'Astrophysique de Paris, F-75005, Paris, France} 
\altaffiltext{6}{Laboratoire AIM-Paris-Saclay, CEA/DSM/Irfu, CNRS, Universit\'{e} Paris Diderot, CE-Saclay, 91191, Gif-sur-Yvette, France}

\email{jwko@kasi.re.kr}

\begin{abstract}

We study the mid-infrared (IR) excess emission of early-type galaxies (ETGs) 
 on the red-sequence at $z <$ 1 using a spectroscopic sample of galaxies 
 in the fields of Great Observatories Origins Deep Survey (GOODS).
In the mass-limited sample of 1025 galaxies with $M_{star}$ $>$ 10$^{10.5}$ $M_{\odot}$ and
 $0.4<z<1.05$, we identify 696 {\it Spitzer} 24 $\mu$m detected (above the 5$\sigma$) galaxies 
 and find them to have a wide range of NUV-$r$ and $r$-[12 $\mu$m] colors 
 despite their red optical $u-r$ colors.
Even in the sample of very massive ETGs on the red sequence with 
 $M_{star}$ $>$ 10$^{11.2}$ $M_{\odot}$, more than 18\% show excess emission 
 over the photospheric emission in the mid-IR.
The combination with the results of red ETGs in the local universe suggests that the recent 
 star formation is not rare among quiescent, red ETGs at least out to $z \sim 1$ if the mid-IR 
 excess emission results from intermediate-age stars or/and from low-level ongoing star 
 formation.
Our color$-$color diagram including near-UV and mid-IR emissions are efficient not only for 
 identifying ETGs with recent star formation, but also for distinguishing quiescent galaxies 
 from dusty star-forming galaxies.

\end{abstract}


\section{INTRODUCTION}

Galaxies in the local universe show a bimodal distribution in the optical color-magnitude
 diagram. 
Quiescent, early-type galaxies (ETGs) populate a narrow 
 red sequence and star-forming, late-type galaxies form a big blue cloud (Strateva et al.
 2001; Blanton et al. 2003; Baldry et al. 2004; Balogh et al. 2004; Choi et al. 2007 
 and references therein). 
This bimodality extends out to at least $z$ $\sim$ 1 (Im et al. 2002; Bell et al. 2004; 
 Willmer et al. 2006; Fritz et al. 2014).
However, the formation and evolution of the red sequence is still not fully understood
 (Faber et al. 2007; Tinker et al. 2013).
One of key populations to address this issue is a transition population 
 that may be in a transition phase migrating into entirely quiescent ellipticals.
In this paper we explore what fraction of red ETGs is in the transition phase 
 and how this fraction evolves since $z \sim$ 1; this provides important constraints 
 on the evolution models for red sequence, especially for massive ETGs.
 
Current galaxy formation models suggest that massive galaxies form most their stars 
 early ($z > 2$; Cowie et al. 1996; Dekel et al. 2009) and are morphologically transformed 
 from spirals into spheroids via major mergers (Khochfar \& Silk 2006; Hopkins et al. 2010).
Thus, most massive quiescent galaxies are already in place at $z$ $\sim$ 1
 (Ilbert et al. 2010, 2013; Moustakas et al. 2013), and ETGs are dominant populations 
 for massive galaxies since $z$ $\sim$ 1 (Buitrago et al. 2013).
Many observations also support that minor (dry) merging is the most likely process to 
 form massive red ETGs at $z$ $<$ 1 because the gas-rich merger rate, 
 important for ETG formation, declines very rapidly since $z$ $\sim$ 1 and 
 their stellar mass growth is very limited since $z$ $\sim$ 1 
 (e.g., Cimatti et al. 2006; Ilbert et al. 2010; Moustakas et al. 2013). 
The minor merger scenario also seems to be one of the most efficient ways to explain a strong
 size evolution of massive red ETGs at least from $z$ $\sim$ 1 (Hopkins et al. 2006; 
 Naab, Johansson \& Ostriker 2009; Newman et al. 2012; Huertas-Company et al. 2013; 
 Damjanov et al. 2014).

Faber et al. (2007) proposed a `mixed' scenario to explain the formation of red ETGs in which
 star formation of blue galaxies is quenched (i.e., shutting off the gas supply) and 
 subsequently they merge further through a series of dry mergers along the red sequence.
Recent studies show that all the red-sequence galaxies are not completely quiescent:
 red ETGs with signs of recent star formation
 (e.g., Yi et al. 2005; Bressan et al. 2006; Schawinski et al. 2007; Clemens et al. 2009; 
 Ko et al. 2009, 2012, 2013; Lee et al. 2010; Vega et al. 2010; Salim et al. 2012); 
 late-type, dust-reddened galaxies with (or without) low-level ongoing star formation 
 (e.g., Im et al. 2002; Wolf, Gray \& Meisenheimer 2005; 
 Bamford et al. 2009; Gallazzi et al. 2009; Wolf et al. 2009; Masters et al. 2010; 
 Ko et al. 2012).
These galaxies in the transition phase can be easily identified if the red-sequence galaxies 
 are examined at different wavelengths. 
The near-ultraviolet (UV) observations reveal a diversity of ETGs depending on 
 the amount of recent ($\leqslant$ 1Gyr) star formation (e.g., Ferreras \& Silk 2000; 
 Kaviraj et al. 2007b). 
The mid-infrared (IR) observations also show that a significant fraction of ETGs
 has excess emission over the photospheric emission (e.g., Bressan et al. 2006; Clemens 
 et al. 2009; Ko et al. 2009, 2012, 2013; Shim et al. 2011; Hwang et al. 2012a).
 
Ko et al. (2013) studied the recent star formation history of local ETGs on the red sequence 
 in the Sloan Digital Sky Survey (SDSS; York et al. 2000).  
We found that among the 648 quiescent red galaxies, 55\% show mid-IR excess emission over 
 the stellar component.
The mid-IR emission seems to mainly originate from the circumstellar dust around 
 asymptotic giant branch (AGB) stars; 
 we excluded the galaxies with active galactic nuclei (AGNs), H$\alpha$ emission, 
 and highly inclined disks.
If we consider only bright ($M_{r}$ $<$ $-$21.5) early-type galaxies,
 the fraction of red galaxies with recent star formation is still 39\%.
We concluded that the recent star formation is common among nearby quiescent, red, 
 early-type galaxies.
This is consistent with the results based on the Galaxy Evolution Explorer 
 ($GALEX$; Martin et al. 2005) that residual star formation is common even for
 bright early-type galaxies (Yi et al. 2005; Schawinski et al. 2007; 
 Kaviraj et al. 2007b; Salim \& Rich 2010).
 
In this study, we use galaxies at $z <$ 1 in the fields of Great Observatories Origins Deep 
 Survey (GOODS; Dickinson et al. 2003; Giavalisco et al. 2004) to investigate the mid-IR
 excess emission of massive ETGs on the red sequence, which is a good indicator of current 
 and/or recent star formation activity ($\sim$1$-$2 Gyrs; Salim et al. 2009; Ko et al. 2013).
Merger (minor) events are speculated to be common at high redshift among ETGs,
 and such events involving gaseous companion galaxies would produce signs of recent star
 formation that can be best caught in the mid-IR.
By studying red ETGs at $z <$ 1, we can test what fraction of them is in the transition phase. 
Section 2 describes the observational data we use. 
We examine the near-UV and mid-IR properties of red ETGs in Section 3, 
 and conclude in Section 4. 
Throughout, we use the AB magnitude system, and adopt flat $\Lambda$CDM 
 cosmological parameters: $H_0 = 70$ km s$^{-1}$ Mpc$^{-1}$, $\Omega_{\Lambda}=0.7$ and 
 $\Omega_{m}=0.3$. 
We also assume a Salpeter initial mass function (Salpeter 1955) for stellar masses and 
 star formation rates (SFR).

\section{THE DATA AND THE SAMPLE}

\subsection{GOODS Sample}

We use a sample of galaxies with spectroscopic redshifts in GOODS.
This sample is originally constructed in Elbaz et al. (2011) and Hwang et al. (2011),
 and here we only give a brief summary of the data set (see also Cervantes-Sodi et al. 2012).

\subsubsection{Multiwavelength Catalog of Galaxies with Spectroscopic Redshifts}

GOODS is a deep multiwavelength survey 
  covering two regions including the Hubble Deep Field North
  and the Chandra Deep Field South. 
Hereafter, we call the two GOODS fields centered on HDF-N and CDF-S
  GOODS-N and GOODS-S, respectively. 
The combined area of the two fields is approximately 320 arcmin$^2$.

Elbaz et al. (2011) made a band-merged catalog of GOODS galaxies
  using the photometric data
  at {\it HST} ACS $BViz$,
  {\it Spitzer} IRAC 3.6, 4.5, 5.8, 8 $\mu$m, 
  IRS peakup array 16 $\mu$m (Teplitz et al. 2011), and 
  MIPS 24 $\mu$m (Magnelli et al. 2011).

Among the sources in the band-merged catalog, 
  we use only 3630 and 3542 galaxies with reliable spectroscopic redshifts 
  over the entire fields of   
  GOODS-N (Cohen et al. 2000; Cowie et al. 2004; Wirth et al. 2004; Reddy et al. 2006; 
  Barger et al. 2008; Cooper et al. 2011) and 
  GOODS-S (Szokoly et al. 2004; Le F{\`e}vre et al. 2004; Mignoli et al. 2005; Vanzella et al.
  2006, 2007, 2008; Ravikumar et al. 2007; Popesso et al. 2009; Kurk et al. 2009; 
  Balestra et al. 2010; Silverman et al. 2010; Xia et al. 2011; Cooper et al. 2012)
  respectively; a typical redshift error is $4 \times 10^{-4}$.

\subsubsection{Clean Index for 24 $\mu$m Detected Galaxies}

The 24 $\mu$m flux densities and their associated errors are determined
  using a PSF fitting technique (Magnelli et al. 2011; Elbaz et al. 2011).  
Because of large PSF size at MIPS 24 $\mu$m sources (FWHM$\sim5.7\arcsec$),
  they use the position of the IRAC 3.6 $\mu$m sources as priors
  for the source extraction at 24 $\mu$m image. 
The advantage of this method is that it can deal with a large part of
  the blending issues in dense fields, 
  and that it can provide a straightforward association of sources at different wavelengths.
This method works well in general.
However, there could be some galaxies in crowded regions whose flux densities
  are not properly deblended.
To deal with these sources separately,
  we compute the level of contamination for each galaxy ($C$), defined by

\begin{equation}\label{eq-con}
C = \sum_{i}\frac{S_{\nu,i}({\rm neib}) e^{-r_{i, {\rm neib}}^2 / (2\sigma^2)}}{S_\nu({\rm target})},
\end{equation} 

where $S_{\nu,i}({\rm neib})$ and $S_{\nu}({\rm target})$ is the flux density of 
 neighboring source and of target, respectively (see also Elbaz et al. 2010; 
 Hwang et al. 2010a).
$r_{i, {\rm neib}}$ is the projected distance between the target and neighbor source, and
 $\sigma = {\rm FWHM}/(2\sqrt{2\rm ln2})$.
We use $S_{\nu}$ at 3.6 $\mu$m, and FWHM at 24 $\mu$m.
We use neighboring sources detected at 3.6 $\mu$m image
  with $r_{i, {\rm neib}}<{\rm FWHM}_{24 \mu m}$.
Note that the neighboring sources are from the photometric catalog 
 regardless of their redshifts.
For ease of computation, we require 
  the neighboring sources to be bright enough to contaminate the
  flux density of target galaxy (i.e., $S_{neib}>0.1S_{target}$).
Figure 1 shows this level of contamination
  of the galaxies with spectroscopic redshifts as a function of 24 $\mu$m flux density.
As expected, most galaxies are distributed around $C=0$;
  there are some galaxies with high $C$ (i.e., might be contaminated).
We call the galaxies with $C\leq0.4$ ``clean'' (93\%) and those with $C>0.4$ ``blended''.
In other words, flux densities of clean galaxies are contaminated less than
 40\%, comparable to the typical error (40\%) of total IR luminosity derived from 
 24 $\mu$m flux density (Elbaz et al. 2010).

\begin{figure}[!ht]
\centering
\includegraphics[width=14cm]{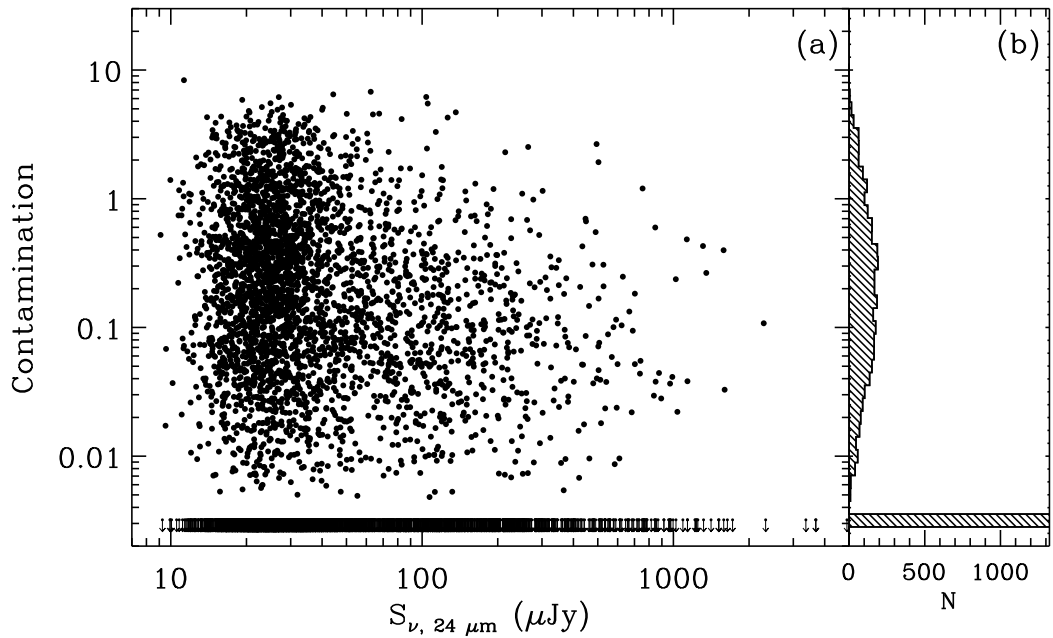} 
\caption{Level of contamination of flux densities by neighbor objects for the GOODS galaxies 
  with spectroscopic redshifts as a function of 24 $\mu$m flux density (a), and 
  its histogram (b).}
\end{figure}

\subsubsection{Mass-limited Sample at 0.4 $ < z <$ 1.05}

Figure 2 shows stellar masses of GOODS galaxies 
  as a function of redshift.
We adopt the stellar mass estimates from Elbaz et al. (2011).
Stellar masses were computed using the photometric data
  from the $U$ band to IRAC 4.5 $\mu$m using the code Z-PEG (Fioc \& Rocca-Volmerange 1999).
The templates used for the stellar mass estimates are
  from a galaxy evolution model P{\'E}GASE.2 (Le Borgne \& Rocca-Volmerange 2002).
The typical error associated with the stellar
  mass estimate is 0.3 dex (see Elbaz et al. 2011 for details).

We define a mass-limited sample of galaxies 
  with $M_{star}$ $\geq$ 10$^{10.5}$ $M_{\odot}$ and 
 0.4 $ \leq z \leq$ 1.05 (enclosed by red solid lines in Figure 2) for further analysis.
We use the mass limit of 10$^{10.5}$ $M_{\odot}$ so that we can have a fair comparison 
 with local ETGs (see Section 2.2).
There are 1025 galaxies in this sample on the MIPS field,
  containing 654 and 371 galaxies for GOODS-N and GOODS-S, respectively.

\begin{figure}[!ht]
\centering
\includegraphics[width=13cm]{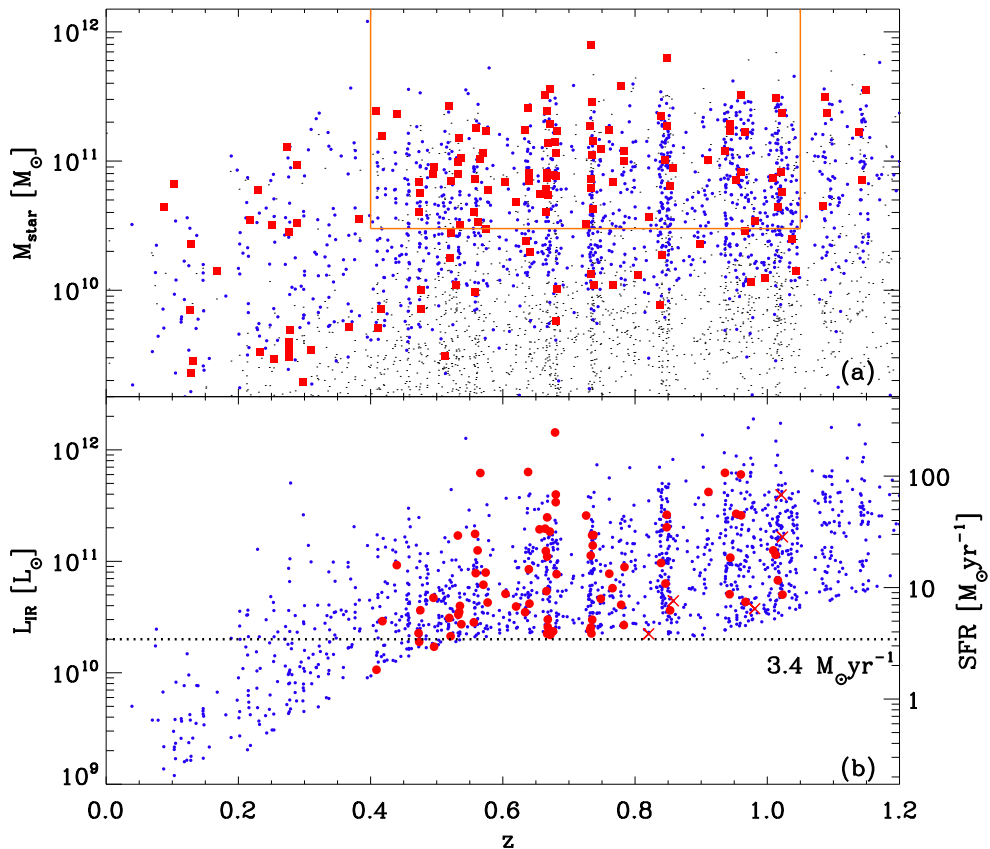}
\caption{$\textit{Upper}$: Stellar mass vs. redshift for the spectroscopic sample of galaxies 
         with $Spitzer$ 24 $\mu$m observation in the GOODS-N and -S (black dots).
         Blue filled circles indicate galaxies with 24 $\mu$m detection (S/N $\geq$ 5), 
         and red filled squares indicate 24 $\mu$m detected, morphologically 
         (see Section 2.1.4.) early-type galaxies.
         Solid lines define a mass-limited sample with $M_{star}$ $\geqslant$ 10$^{10.5}$ 
         $M_{\odot}$ and 0.4 $\leq$ $z$ $\leq$ 1.05.
         $\textit{Lower}$: IR luminosity vs. redshift for 24 $\mu$m detected galaxies 
         (blue filled circles are the same as above).
         Red filled circles and crosses denote ``clean'' and ``blended'' 24 $\mu$m detected,
         early-type galaxies in the mass-limited sample, respectively.
         The right axis is the SFR converted from the IR luminosity using the relation
         in Kennicutt (1998):
         SFR ($M_{\odot}yr^{-1}$) = 1.72$\times$10$^{-10}$ $L_{IR}/L_{\odot}$.
        }
\end{figure}

The lower panel of Figure 2 shows total IR luminosities, $L_{IR}$, as a function of redshift 
 for 24 $\mu$m detected galaxies.
We compute the total IR luminosity ($L_{IR}$) from the 24 $\mu$m flux density using the SED 
 templates of Chary \& Elbaz (2001). 
Elbaz et al. (2010) found that IR luminosities from the 24 $\mu$m flux densities agree well
 with those from $Herschel$ PACS and SPIRE far-IR data, with a dispersion $<$ 0.15 dex. 
They also found that extrapolations from 24 $\mu$m measurements work well up to $z \sim 1.3$.
The $Spitzer$ 24 $\mu$m detection limit ($\sim$30 $\mu$Jy, 5$\sigma$; Magnelli et al. 
 2009) in both GOODS fields is quite flat at 0.4 $ \leq z \leq$ 1.0, and corresponds to
 $L_{IR}$ $\sim$ $2\times$ 10$^{10}$ $L_{\odot}$. 
This is comparable to a SFR of $\sim$3.4 $M_{\odot}yr^{-1}$ using the Kennicutt (1998) 
 relation.


We compute the rest-frame $r$-band absolute magnitudes ($M_{r}$), and the 
 rest-frame (NUV$-r$), ($u-r$), and $r-$[12 $\mu$m] colors using the $Le Phare$ code 
 (Arnouts et al. 1999; Ilbert et al. 2006). 
We use all the data available from $HST$ ACS $B$ to $Spitzer$ 24 $\mu$m for the SED fit. 
We use the SED templates generated with BC03 model (Bruzual \& Charlot 2003) by assuming 
 an exponentially declining star formation history SFR $\propto$ $e^{-t/\tau}$ 
 ($\tau$ between 0.1 Gyr to 30 Gyr).
The SEDs were generated for a grid of 44 ages (0.1 Gyr to 13.5 Gyr) and three different 
 metallicities (Z=0.02, 0.008, and 0.004), and dust extinction was added using 
 the formula of Calzetti et al. (2000) for $E_{B-V}$ between 0 to 0.5.
For IR galaxy SEDs, we use Chary \& Elbaz (2001) for 16 and 24 $\mu$m flux densities.
As a sanity check, we also compute $r-$[12 $\mu$m] using the SED templates of 
 Assef et al. (2010) for all the photometric data. 
The $r-$[12 $\mu$m] colors from these two methods agree with a dispersion $<$ 0.2 mag, 
 much smaller than the spread of $r-$[12 $\mu$m] colors of ETGs on the red sequence 
 (see Fig. 6).

\subsubsection{Early-type Galaxies on the Red Sequence}

Because we are interested in the mid-IR activity of early-type galaxies,
  we need to identify morphologically early-type galaxies
  in the mass-limited sample of galaxies in Figure 2.
To do that, we adopt galaxy morphology data from Hwang \& Park (2009) and Hwang et al. (2011)
  that is based on the visual inspection of ACS $BViz$ images.
We classify the galaxies into two groups: 
  early types (E/S0) and late types (S/Irr).
Early-type galaxies are those with little fluctuation in the surface brightness
 and color and with good symmetry, but late-type galaxies show internal structures 
 and/or color variations in the pseudocolor images.
This classification agrees very well ($>98\%$) with those of Bundy et al. (2005)
  who also performed a visual morphological classification for GOODS galaxies.
We perform additional visual classification for the galaxies in the mass-limited sample
  of galaxies that are not included in Hwang \& Park (2009) and Hwang et al. (2011).
We also verify the morphological classification of the galaxies
  covered by the Cosmic Assembly Near-Infrared Dark Energy Legacy Survey
  with the NIR {\it HST} Wide Field Camera 3 images (Grogin et al. 2011; 
  Koekemoer et al. 2011).
Figure 3 shows examples of
  {\it HST} $zH$ and {\it Spitzer} 24 $\mu$m cutout images
  for early-type galaxies in GOODS-N and -S.
  From top to bottom, 
  they are those without 24 $\mu$m detection in GOODS-N and -S, 
  those with 24 $\mu$m detection in GOODS-N and -S, and
  those with 24 $\mu$m detection but blended in GOODS-N and -S.

\begin{figure}
\centering
\includegraphics[width=13cm]{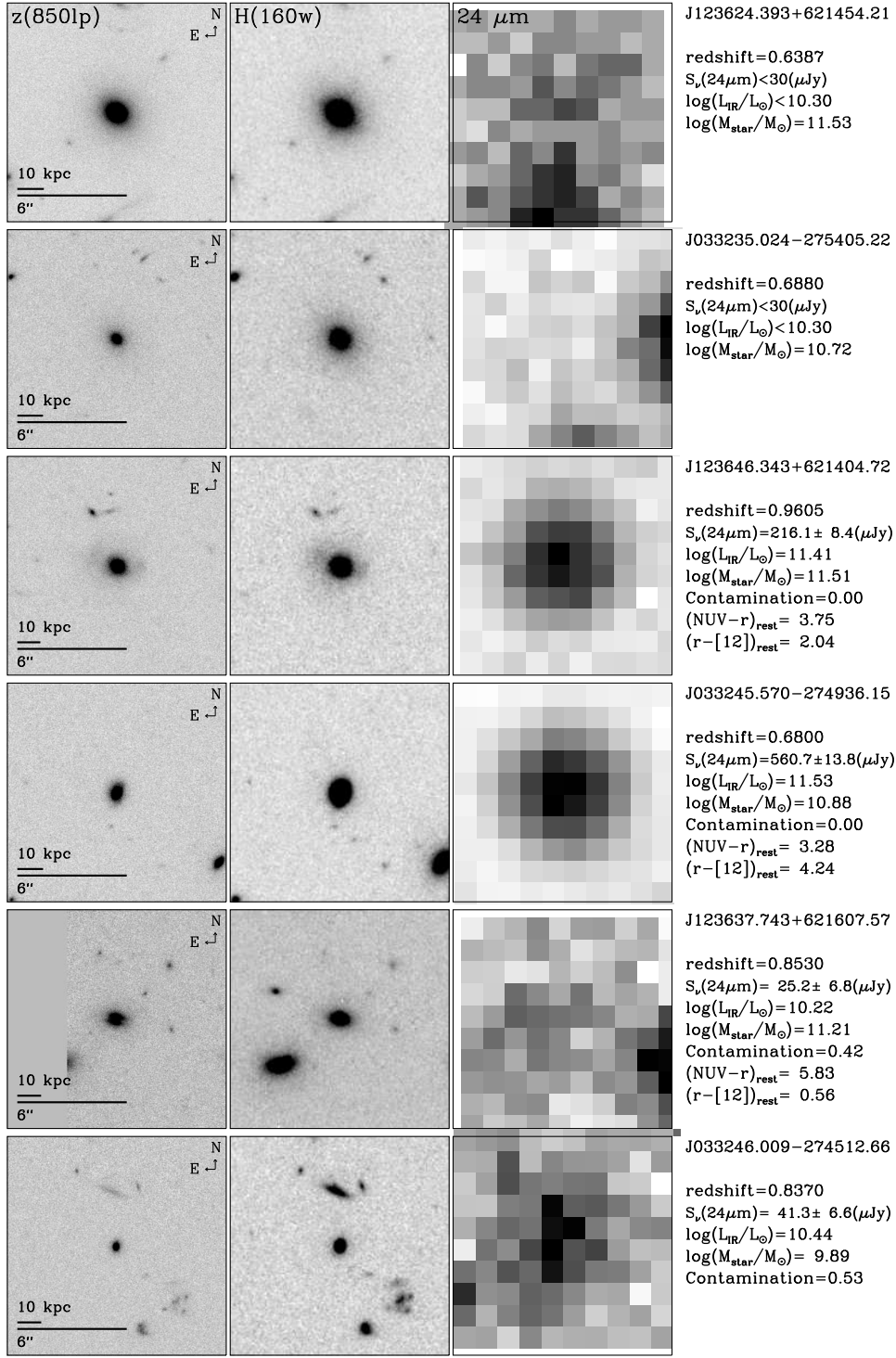} 
\caption{Example {\it HST} $zH$ and {\it Spitzer} 24 $\mu$m cutout images
  ($12\arcsec \times 12\arcsec$) for early-type galaxies. 
  From top to bottom: those without 24 $\mu$m detection 
    in GOODS-N (J123624.393$+$621454.21) and
    in GOODS-S (J0333235.024$-$275405.22),
    those with 24 $\mu$m detection 
    in GOODS-N (J123646.343$+$621404.72) and
    in GOODS-S (J033245.570$-$274936.15), and 
    those with 24 $\mu$m detection but blended
    in GOODS-N (J123637.743$+$621607.57) and
    in GOODS-S (J033246.009$-$274512.66).
    Because there are no blended, 24 $\mu$m detected (S/N $>$ 5) galaxies with $HST$ 
    $H$-band images in the mass-limited sample, in the bottom panels, we plot the 
    galaxies with S/N $<$ 5 at 24 $\mu$m or $M_{star} $ $<$ 10$^{10.5}M_{\odot}$.     
    }
\end{figure}

We use rest-frame $u-r$ color versus $r$-band absolute magnitude ($M_{r}$) 
 to identify red-sequence galaxies at $z < 1$.
In Figure 4, we show the distribution of rest-frame $u-r$ color against $M_{r}$ 
 in three redshift bins.  
Solid lines indicate a fit to the color-magnitude relation (CMR) for red-sequence galaxies 
 with $-$0.08 that is derived from local SDSS galaxies (Ko et al. 2013; see Figure 4 therein).   
This slope is similar to SDSS luminous red galaxies (Baldry et al. 2004). 
This slope is also compatible with the slope of $-$0.07 (the standard deviation of 
 residuals to the fit of $\sim$0.2 mag) derived from GOODS ETGs with $M_{r}$ $<$ $-$21 
 and 0.4 $ \leq z \leq$ 1.05. 
The red-sequence galaxies become redder by $\sim$0.1 mag from the highest redshift bin to
 the lowest redshift bin.
Dashed lines indicate the color cut adopted in this study to separate red and blue galaxies.
The CMR is moved to blueward of the linear fit by 0.25 mag to define the color cut.
The amount of offset (i.e. 0.25 mag) is a standard deviation from a fit to the histogram of
 $u-r$ colors for massive ($M_{star} $ $>$ 10$^{10.5}M_{\odot}$) galaxies
 regardless of 24 $\mu$m detection at 0.4 $\leq z \leq$ 1.05.

In the result, in the mass-limited sample of 1025 galaxies (defined in Fig. 2; 32\% are ETGs)
 68\% (696) is detected at 24 $\mu$m (blue small circles in Fig. 4). 
Among the 696 24 $\mu$m detected galaxies, 82 galaxies are ETGs (12\%, red filled circles 
 in Fig. 4) that we use for the following analysis. 
Among the 82 ETGs, 85\% (70) are on the red sequence. 
Note that there are only five galaxies out of 82 ETGs flagged as ``blended" (crosses 
 in Fig. 2); their 24 $\mu$m flux densities could be overestimated.


To remove the contribution of dust heated by AGN, we identify AGN-host galaxies using 
 $Spitzer$ IRAC data (e.g., Lacy et al. 2004; Stern et al. 2005; Assef et al. 2010; 
 Jarrett et al. 2011; Lee et al. 2012). 
There are only four galaxies that satisfy the AGN selection criteria of Stern et al. (2005),
 thus AGN contamination is negligible in the following analysis.

\begin{figure}[!ht]
\centering
\includegraphics[width=12cm]{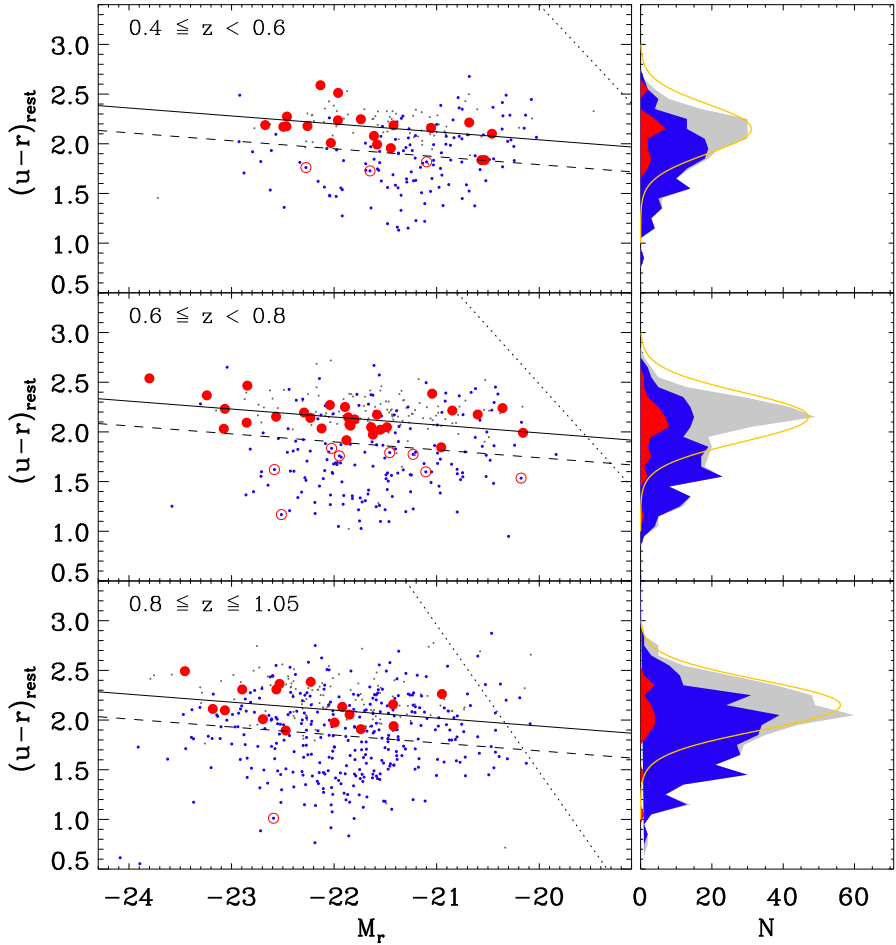}
\caption{Rest-frame $u-r$ colors vs. $r$-band absolute magnitudes for the spectroscopic 
         sample of GOODS galaxies with $M_{star}$ $>$ 10$^{10.5}$ $M_{\odot}$ 
         (definded in Fig. 2; gray dots).
         Blue filled circles represent the galaxies with 24 $\mu$m 
         detection. Red filled and open circles 
         show ``clean" red and blue ETGs with 24 $\mu$m detection, 
         respectively.
         Solid lines are the best-fits of the color-magnitude relation for 
         red-sequence galaxies with a fixed slope of $-$0.08.         
         Dashed line indicates a cut for selecting red-sequence galaxies, 
         moved from the best-fit color-magnitude relation to a bluer color by 0.25 mag.       
         Slant dotted lines at faint ends indicate a limiting magnitude for the 
         spectroscopic sample, $i_{F775W}$ $\leq$ 23.5.
         The right panels show the distribution of
         rest-frame $u-r$ color for each sample in the left panels. 
         Yellow curves are the best-fit Gaussians ($\sigma$ of 0.25 and center of 2.15) 
         to the histograms of the galaxies with 
         $M_{star}$ $>$ 10$^{10.5}$ $M_{\odot}$ and 0.4 $\leq$ $z$ $\leq$ 1.05. 
          }
\end{figure}

\subsection{SDSS Sample}

To compare the results of GOODS galaxies at high redshift with those in the local universe,
 we use the spectroscopic sample of galaxies in the SDSS Data Release 7 
 (DR7, Abazajian et al. 2009). 
We also use a photometric sample of SDSS galaxies whose redshift information is not 
 available in the SDSS database, but available in the literature (Hwang et al. 2010b).
The stellar mass estimates are from MPA/JHU DR7 value-added galaxy
 catalog (VAGC\footnote{http://www.mpa-garching.mpg.de/SDSS/DR7/}); 
 these are computed using SDSS five-band photometric data with the model of 
 Bruzual \& Charlot (2003) (Kauffmann et al. 2003; Gallazzi et al. 2005).
We adopt galaxy morphology data from the Korea Institute for Advanced Study (KIAS) DR7
 VAGC\footnote{http://astro.kias.re.kr/vagc/dr7/} (Choi et al. 2010);
 these are divided into two morphological types, the same criteria as for GOODS galaxies.

We identify the mid-IR counterparts of the SDSS galaxies in the Wide-field
 Infrared Survey Explorer ($WISE$) source catalog (see Ko et al. 2013 for details).
Figure 5 shows stellar masses (a) and total IR luminosities (b) of these SDSS galaxies with
 $WISE$ 22 $\mu$m detection as a function of redshift.  
We define a mass-limited sample at 0.04 $\leq$ $z$ $\leq$ 0.11 and 
 $M_{star}$ $>$ 10$^{10.5}$ $M_{\odot}$, the same mass limit as for GOODS galaxies. 
By adopting a method similar to the one applied to GOODS galaxies, we compute the IR 
 luminosities for the SDSS galaxies from the 22 $\mu$m flux densities. 
The SFRs converted from these 22 $\mu$m-derived IR luminosities are consistent with those
 based on optical emission lines (Hwang et al. 2012a,b; Lee et al. 2013).
The dashed line in Figure 5(b) represents $\sim$2 $\times$ 10$^{10}$ $L_{\odot}$,
 corresponding to $\sim$3.4 $M_{\odot}yr^{-1}$.

\begin{figure}[!ht]
\centering
\includegraphics[width=13cm]{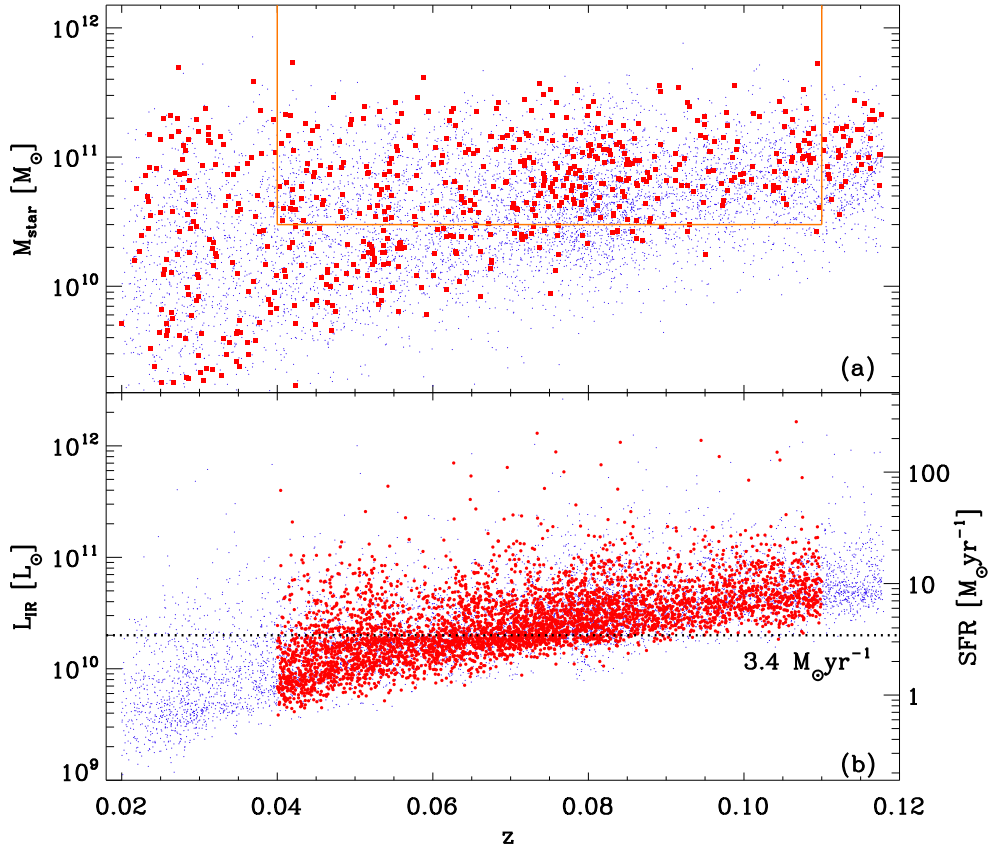}
\caption{$\textit{Upper}$: Same as Fig. 2(a), but for SDSS galaxies.
         Blue dots denote $WISE$ 22 $\mu$m detected galaxies with S/Ns $\geq$ 5,
         and red filled squares indicate 22 $\mu$m detected, morphologically ETGs
         (only 10\% of galaxies randomly selected in the total sample are shown).
         Solid lines define the mass-limited sample at 0.04 $\leq$ $z$ $\leq$ 0.11 and
          $M_{star}$ $\geqslant$ 10$^{10.5}$ $M_{\odot}$.
         $\textit{Lower}$: Same as Fig. 2(b), but for SDSS galaxies. 
         Red filled circles indicate $WISE$ 22 $\mu$m detected,
         ETGs in the mass-limited sample.        
         The right axis is the SFR converted from the IR luminosity using the relation 
         in Kennicutt (1998):
         SFR ($M_{\odot}yr^{-1}$) = 1.72$\times$10$^{-10}$ $L_{IR}/L_{\odot}$.}
\end{figure}

\section{RESULTS AND DISCUSSION}

\subsection{Near-UV and Mid-IR Excess emissions of Red Early-type Galaxies}

Here we examine near-UV and mid-IR emissions of red ETGs in GOODS. 
We first show how local galaxies and SED templates populate in near-UV/optical/mid-IR 
 color domain (Section 3.1.1), and then discuss how high-z GOODS galaxies are distributed 
 in this domain (Section 3.1.2).

\begin{figure}[!ht]
\centering
\includegraphics[width=12cm]{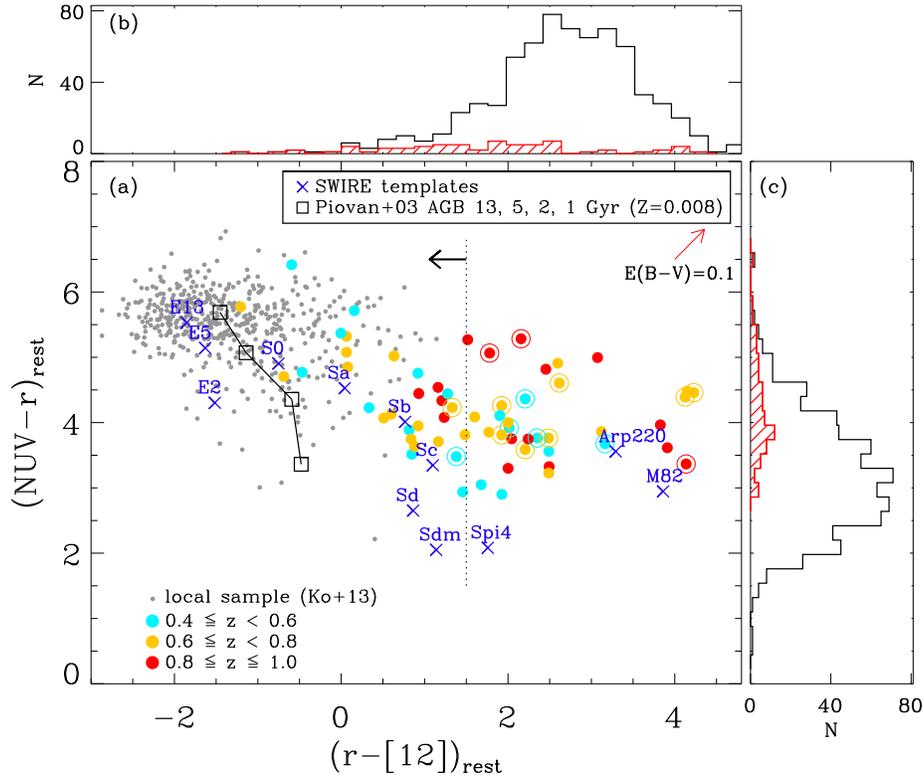}
\caption{Rest-frame $r-$[12 $\mu$m] vs. NUV$-r$ colors for 
         $Spitzer$ 24 $\mu$m detected GOODS red ETGs with $M_{star} (M_{\odot})$ $>$
         10$^{10.5}$ (large filled circles color-coded by their redshifts).
         Dotted line indicates the upper limit for $Spitzer$ 24 $\mu$m flux densities
         (5$\sigma$).
         Large open circles represent $Herschel$ PACS-100 $\mu$m detection (above the
         3$\sigma$ limit).
         For comparison, gray dots indicate $WISE$ 12 $\mu$m 
         detected quiescent, red ETGs with $M_{star} (M_{\odot})$ $>$ 10$^{10.5}$ 
         and 0.04 $\leq$ $z$ $\leq$ 0.11 (Ko et al. 2013).
         We show the extinction-free rest-frame colors of the SWIRE templates of 
         Polletta et al. (2007) including three ellipticals (2, 5, and 13 Gyr), 
         seven spirals (S0, Sa, Sb, Sc, Sd, Sdm, and Spi4), and two starbursts 
         (M82 and Arp220).
         We also plot the SSP AGB models (Piovan et al. 2003) along mean stellar ages 
         (1, 2, 5, and 13 Gyr) for 40\% solar metallicity (Z=0.008).
         Open and hatched histograms denote rest-frame $r-$[12 $\mu$m] (b) and NUV$-r$ (c)
         colors for $Spitzer$ 24 $\mu$m detected massive ($M_{star} (M_{\odot})$ 
         $>$ 10$^{10.5}$) galaxies and red ETGs, respectively. 
         }
         
\end{figure}

\begin{figure}[!ht]
\centering
\includegraphics[width=12cm]{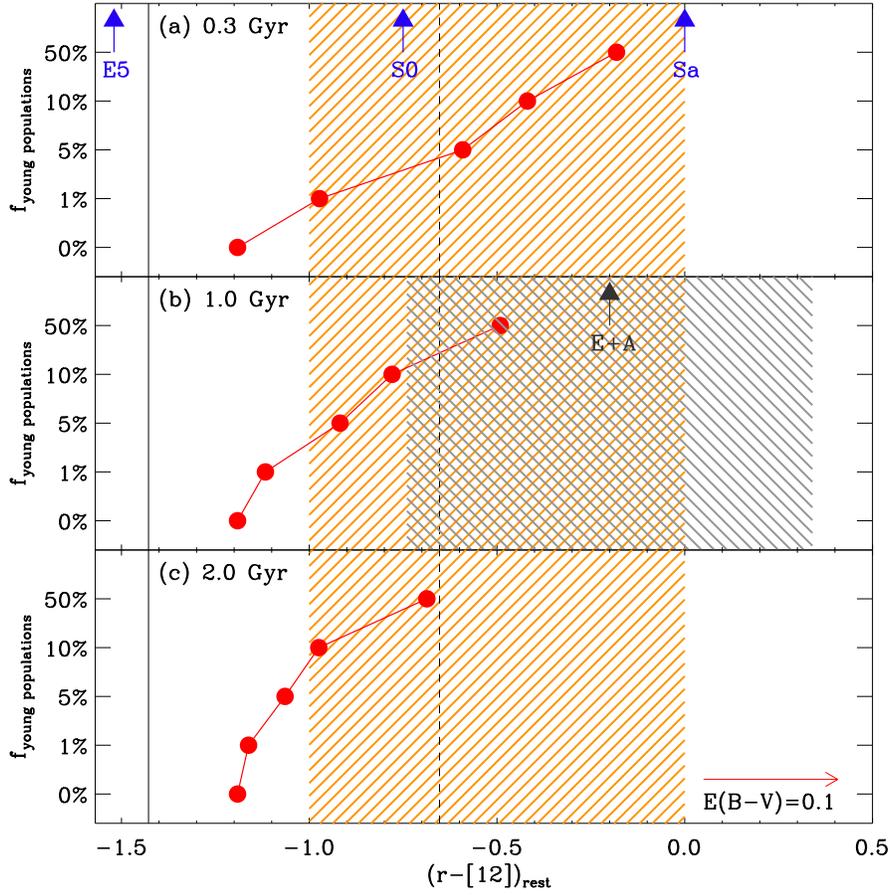}
\caption{Two component SSP model $r-$[12 $\mu$m] colors are composed of young populations
         (0.3, 1.0 and 2.0 Gyr) with fractions of 1, 5, 10 and 50\%, 
         and an old (10 Gyr) underlying population with a solar metallicity.  
         $r-$[12 $\mu$m] colors of the SWIRE templates E5, S0 and Sa are also overplotted 
         with blue arrows.
         Vertical solid and dashed lines in all panels represent the medain value and its
         standard deviation of local red ETGs 
         with $M_{star} (M_{\odot})$ $>$ 10$^{10.5}$ (gray dots in Figure 6), respectively.
         In the middle panel (b), upward arrow and gray hatched ($ \diagdown\diagdown$) 
         represent the median value and its standard deviation of early-type E+A 
         galaxies with $M_{star} (M_{\odot})$ $>$ 10$^{10.5}$ from Choi et al. (2009),
         respectively. 
         Orange hatched ($ \diagup\diagup$) in all panels indicate the region of
         quiescent red ETGs with mid-IR excess defined in Section 3.2.1. 
         }
         
\end{figure}

\subsubsection{At Low Redshifts: SDSS Galaxies}

The optical CMR is commonly used to separate red ETGs from blue late-type 
 galaxies out to $z$ $\sim$ 1 (e.g., Kodama et al. 1999; Bell et al. 2004; Faber et al. 2007). 
However, many studies revealed that nearby, red ETGs show a large scatter in the 
 near-UV$-$optical color-magnitude diagrams with signs of recent ($\leq1$ Gyr) star formation.
Furthermore, Ko et al. (2013) showed that a half of nearby, bright, red ETGs
 with $WISE$ 12 $\mu$m detection show mid-IR excess emission over the stellar component.
The mid-IR emission can have much larger contribution from intermediate-age stars 
 than from young stars (Salim et al. 2009; Kelson \& Holden 2010; Donoso et al. 2012), and
 can give us an unobscured view of the star formation activity.
We found that most quiescent (without optical emission lines) red ETGs with 
 near-UV excess show mid-IR excess emission, while less than 50\% of 
 red ETGs with mid-IR excess show near-UV excess emission (Ko et al. 2013). 
These support the idea that the mid-IR emission is sensitive to star 
 formation with longer timescales ($\geqslant$ 1 Gyr) than near-UV emission.

In Figure 6(a), gray dots indicate $WISE$ 12 $\mu$m detected local 
 (0.04 $\leq$ $z$ $\leq$ 0.11) red ETGs with $M_{star}$ $>$ 10$^{10.5}$ $M_{\odot}$
 (Ko et al. 2013).
In this local sample, we exclude galaxies with H$\alpha$ emission, AGN features, 
 and highly inclined disks.
For reference, we show the extinction-free rest-frame colors of the SWIRE templates of 
 Polletta et al. (2007) including three ellipticals (2, 5, 13 Gyr), 
 seven spirals (S0, Sa, Sb, Sc, Sd, Sdm, Spi4), and two starbursts (M82, Arp220).
These templates are generated with the GRASIL code (Silva et al. 1998) 
 in which the effects of dusty envelopes around AGB stars are taken into account by adopting
 the procedure of Bressan et al. (1998). 
We also overplot the predictions from Single Stellar Population (SSP) models that include 
 the emission from circumstellar dust around AGB stars (solid line; Piovan et al. 2003),
 along mean stellar ages (1, 2, 5, and 13 Gyr) for 40\% solar metallicity.
Although two models predict $r-$[12 $\mu$m] colors differently, 
Ko et al. (2013) showed that redder $r-$[12 $\mu$m] colors (mid-IR excess) and 
 bluer NUV$-r$ colors (near-UV excess) of local red ETGs seem to be consistent 
 with the two-component SSP model predictions.

To roughly estimate a contribution of young populations to these mid-IR detected red ETGs, 
 we plot a stellar mass fraction of young populations as a function of $r-$[12 $\mu$m]   
 colors in Figure 7. 
We use two-component SSP models; we assume that an old underlying population
 (10 Gyr; an instantaneous burst of star formation at $z$ $\sim$ 3)
 with a solar metallicity and young populations (0.3, 1.0 and 2 Gyr) with fractions of 
 1, 5, 10 and 50\%. 
For comparison, we also mark $r-$[12 $\mu$m] colors of the SWIRE templates of E5, S0 and Sa
 with arrows.
Vertical solid and dashed lines in all panels represent the medain value and its
 standard deviation of local red ETGs with $M_{star} (M_{\odot})$ $>$ 10$^{10.5}$ 
 (gray dots in Figure 6), respectively.
The $r-$[12 $\mu$m] color of E5 template corresponds to the median value of local red ETGs
 and the color of Sa template roughly coincides with 
 an sSFR cut (log sSFR = $-$10.7) to separate quiescent galaxies
 from star-forming ones (see Section 3.2.1 for details).
Orange hatched regions in all panels indicate quiescent red ETGs with mid-IR excess 
 defined in Section 3.2.1. 
This shows that only $\sim$1\% of 0.3 Gyr, $\sim$5\% of 1 Gyr, and  
 $\sim$10\% of 2 Gyr young populations can change $r-$[12 $\mu$m] colors to move the galaxies 
 into the mid-IR excess region, indicating that mid-IR is likely to trace
 star formation over longer (up to $\sim$2 Gyr) timescales.
It should be noted that more than 10\% of 0.3 Gyr make $u-r$ color bluer than our red
 sequence selection criterion, thus only $<$ 5\% with dust extinction can explain the 
 $r-$[12 $\mu$m] color distribution of our local red ETGs.
For comparison, the median value of $r-$[12 $\mu$m] colors for local E+A galaxies 
 (sign of recent starburst within $\sim$1 Gyr) from Choi et al. (2009) is shown
 in the middle panel of Figure 7. 
Among the E+A sample, we use only morphologically early-type galaxies 
 with $M_{star} (M_{\odot})$ $>$ 10$^{10.5}$.
At least 10\% and more than 50\% of 1 Gyr populations with dust extinction can 
 produce the color distribution of post-starburst systems, consistent 
 with previous results (e.g., Kaviraj et al. 2007a).

\subsubsection{At High Redshifts: GOODS Galaxies}

In Figure 6(a), we also plot 24 $\mu$m detected, GOODS red ETGs at 
 0.4 $\leq$ $z$ $\leq$ 1.05 and $M_{star} (M_{\odot})$ $>$ 10$^{10.5}$ (large circles).
The upper limit of $r-$[12 $\mu$m] colors from the 24 $\mu$m detection limit ($<5\sigma$) 
 is indicated with vertical dotted line and arrow (i.e. $r-$[12 $\mu$m] $<$ 1.5)
 \footnote{This color cut is insensitive to redshift because the $r-$[12 $\mu$m] color is
 a good proxy of sSFR, and the detection limit of $L_{IR}$ (i.e., SFR) computed with 
 the 24 $\mu$m flux density is quite flat for the mass-limited sample 
 at 0.4 $\leq$ $z$ $\leq$ 1.05 in this study.}. 
We also mark the galaxies detected ($>$ $3\sigma$) at $Herschel$ 100 $\mu$m band with 
 large open circles.

GOODS red ETGs are expected to be around 3 Gyr SSP model if we assume that they are 
 single-burst stellar populations formed at $z=$ 3 so that their age is $\sim$ 3 Gyr 
 at $z=$ 1.
The 24 $\mu$m undetected galaxies have $r-$[12 $\mu$m] $<$ 1.5, suggesting that they have 
 an old (3 Gyr) underlying population with young and intermediate-age ($\leq$ 2 Gyr)
 populations. 
Most of these 24 $\mu$m undetected galaxies are ETGs (76\%, 250 out of 329), 
 and are on the red sequence (see right panels of Fig. 4).
However, we note that red ETGs with $r-$[12 $\mu$m] $>$ 0 seem to be
 dust-reddened or/and have low-level ongoing star formation because two-component SSPs
 can not make their red colors (see Fig. 7).
  
On the other hand, more than half of 24 $\mu$m detected GOODS red ETGs have 
 $r-$[12 $\mu$m] $>$ 1. 
Their colors are consistent with Spiral galaxy templates with dust extinction. 
Considering that Sc is more actively forming stars than Sa, their optical red colors suggest 
 that Sa galaxies are weakly forming stars, not enough to change their integrated optical
 colors (similar to red spirals in Galaxy Zoo of Masters et al. 2010 and to optically passive
 spirals in Wolf et al. 2009). 
 However, Sc galaxies are normal star-forming galaxies, but suffer from heavy dust extinction. 
Because our morphology classification is not sensitive to separate Sa from E/S0, there could
 be some Sa galaxies classified as ETGs in this study. 
Interestingly, about 41\% (14/34) of galaxies with $r-$[12 $\mu$m] $>$ 1.5 are detected at 
 PACS 100 $\mu$m, suggesting that some of 24 $\mu$m detected, red ETGs are currently forming
 stars despite their early-type morphology (see Fukugita et al. 2004; Lee et al. 2006, 2010;
 Skibba et al. 2011; Hwang et al. 2012a; Ko et al. 2012; Martini et al. 2013).
 
In summary, while most GOODS ETGs remain red in the $u-r$ colors, 
 they show a wide range of NUV$-r$ colors (i.e., near-UV excess), indicating 
 that a large fraction of red ETGs at $z$ $<$ 1.0 have experienced star formation within 
 $\sim$1 Gyr or/and are undergoing low-level star formation. 
This is because the near-UV emission from young massive stars is more sensitive to 
 recent star formation than the blue-band optical emission even though the contribution of
 young massive stars to the total stellar mass is only 1\%$-$3\% (Kaviraj et al. 2007b).
Similarly, a wide range of $r-$[12 $\mu$m] colors seems to show a variety of 
 mid-IR emission for GOODS red ETGs.
Here the mid-IR excess emission of galaxies may attribute to intermediate-age stars 
 (i.e., sensitive to star formation over relatively longer timescales of $\sim$2 Gyrs) or/and
 to low-level ongoing star foramtion.
In addition, red ETGs with reddest $r-$[12 $\mu$m] colors show strong mid-IR excess 
 due to dust, thus their sSFRs are similar to normal star-forming galaxies.
Therefore, near-UV$-$optical versus optical$-$mid-IR color-color diagram is 
 very efficient for detecting ETGs with recent (within 2 Gyrs) star formation and 
 breaking the degeneracies between dusty star-forming and quiescent galaxies.

\subsection{Specific SFRs of Red Early-type Galaxies}

In the local universe, low-level recent star formation of ETGs has been detected through
 near-UV excess emission (e.g., Yi et al. 2005; Salim \& Rich 2010) and mid-IR excess
 emission (e.g., Bressan et al. 2006; Ko et al. 2013).
With an observational result that there is no significant evolution in the optical CMR 
 out to, at least, $z = 1$ (e.g., Im et al. 2002; Bell et al. 2004; Willmer et al. 2006; 
 see also Fig. 4),
 the evolution of red ETGs with recent star formation can provide important constraints 
 on the formation models of present-day massive red ETGs.
In this section, we first study the change in specific SFR (sSFR = SFR$_{IR}/M_{star}$)
 of red ETGs with redshift (Section 3.2.1), and then compute the fraction of red ETGs with
 mid-IR excess emission in Section 3.2.2.

\subsubsection{Stellar Mass Dependence}

In the following, we assume that the mid-IR excess emission of red ETGs is dominated by 
 intermediate-age stars or/and low-level current star formation.
We adopt the specific SFR to trace this excess emission by 
 taking into account an increase in the average SFRs of star-forming galaxies with
 redshift in Elbaz et al. (2011).
Thus, the sSFR cut of ETGs with mid-IR excess emission also changes with
  redshift; we use the sSFR cut used for local ETGs with mid-IR excess emission as a
  reference.

\begin{figure}[!ht]
\centering
\includegraphics[width=16cm]{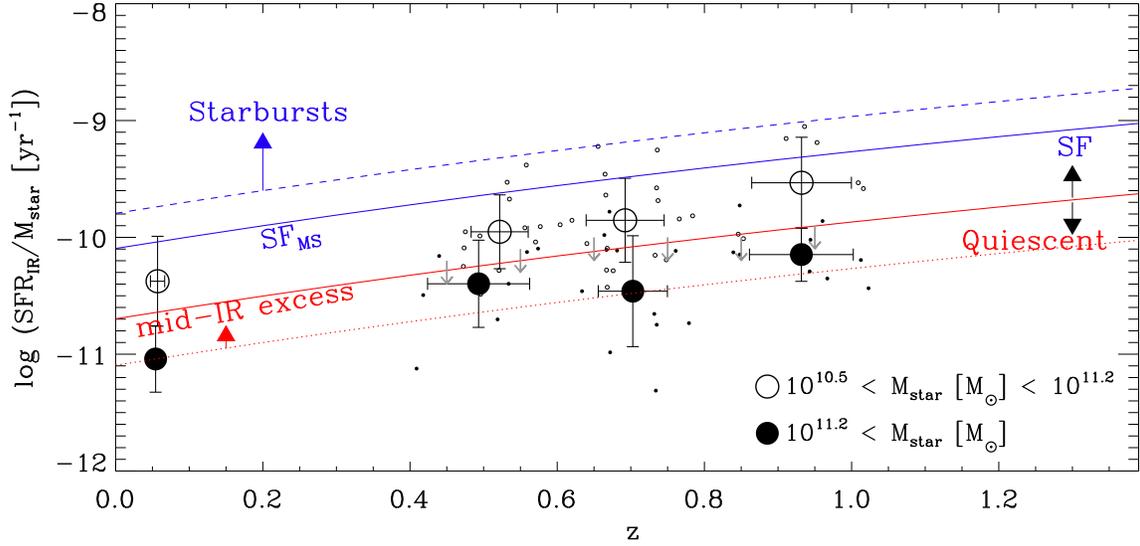}
\caption{Specific SFR (sSFR = SFR$_{IR}/M_{star}$) vs. redshift for $Spitzer$ 24 $\mu$m
         detected GOODS and $WISE$ 22 $\mu$m detected SDSS red ETGs with 
         $M_{star}$ $>$ 10$^{10.5}$ $M_{\odot}$.
         We divide the sample to two groups based on their stellar masses:
         small filled and open circles represent 
         high-mass ($M_{star}$ $>$ 10$^{11.2}$ $M_{\odot}$) and 
         low-mass ($M_{star}$ $<$ 10$^{11.2}$ $M_{\odot}$) ones, respectively.
         Large filled and open circles are median values at each bin and
         error bars denote standard deviations.
         Blue solid line is the mean evolutionary trend of IR-selected star-forming
         galaxies (i.e. main sequence galaxies, see eq. 13 and Fig. 18 in Elbaz et al. 2011).
         Blue dashed line is for separating starburst galaxies from main sequence galaxies
         (see eq. 14 in Elbaz et al. 2011).          
         The red solid line is an sSFR cut for separating star-forming galaxies from 
         quiescent ones (see Figure 6 of Ilbert et al. 2010), which is smaller than 
         the mean evolutionary trend of main-sequence galaxies by a factor of 4 
         (log sSFR = $-$10.7 at $z$ = 0, see Fig. 6 in Ilbert et al. 2010). 
         Red dotted line is an sSFR cut to distinguish quiescent galaxies from mid-IR excess
         galaxies (defined by 0.1$\times$sSFR$_{MS}$; see also Fig. 15 in Ko et al. 2012). 
         Small gray downward arrows indicate the detection limit of 24 $\mu$m flux density 
         for the mass-limited sample.}
         
\end{figure}

In Figure 8, we plot sSFRs of $Spitzer$ 24 $\mu$m detected GOODS red ETGs and of $WISE$ 22
 $\mu$m detected SDSS red ETGs with $M_{star}$ $>$ 10$^{10.5}$ $M_{\odot}$
 as a function of redshift.   
We further divide the galaxy sample into two groups based on their stellar masses;
 large circles show median values of each bin. 
More massive red ETGs (filled circles) certainly have lower sSFRs values at $z < 1$
	\footnote{The limit for sSFR is lower for more massive galaxies than less massive ones, 
	but there are few or no red ETGs with $M_{star}$ $>$ 10$^{11.2}$ $M_{\odot}$ 
	above the sSFR limit (small gray downward arrows in Fig. 8) for the mass-limited	
	sample ($M_{star}$ $>$ 10$^{10.5}$ $M_{\odot}$).}. 
Rodighiero et al. (2010) and Karim et al. (2011) also found that more massive 
 galaxies have lower sSFRs in this redshift range, using $Herschel$ PACS far-IR data 
 and stacked 1.4 GHz radio continuum emission data, respectively.

The blue solid line in Figure 8 is the mean evolutionary trend of IR-selected star-forming
 galaxies (i.e. main sequence galaxies, see eq. 13 and Fig. 18 in Elbaz et al. 2011). 
The blue dashed line is for separating starburst galaxies from main sequence galaxies 
 (see eq. 14 in Elbaz et al. 2011). 
It is interesting to see that the median sSFRs for both mid-IR detected GOODS and SDSS ETGs 
 with  $M_{star}$ $>$ 10$^{9.5}$ $M_{\odot}$ are consistent with main sequence galaxies. 
The red solid line is a factor of 4 below the main sequence (blue solid line), 
 indicating an sSFR cut (log sSFR = $-$10.7 at $z$ $\sim$ 0) to separate
 quiescent galaxies from star-forming ones (Gallazzi et al. 2009; Ilbert et al. 2010; 
 Ko et al. 2012).
Small, gray downward arrows indicate the detection limit of $Spitzer$ 24 $\mu$m flux density
 for our mass-limited ($M_{star}$ $>$ 10$^{10.5}$ $M_{\odot}$) sample.
All of them are below or around the sSFR cut for quiescent galaxies, indicating that 24 $\mu$m
 undetected galaxies at 0.4 $<$ $z$ $<$ 1.0 are quiescent systems.
Galaxies with mid-IR excess are defined as those with an sSFR $>$ 0.1 $\times$ sSFR$_{MS}$, 
 corresponding to the sSFR cut to select quiescent red-sequence galaxies with 
 mid-IR excess at $z$ $\sim$ 0.09 (red dotted line; see Figure 15 of Ko et al. 2012). 
 
Figure 8 shows that less massive ETGs are more likely to have mid-IR excess emission than 
 more massive ETGs at $z <$ 1, consistent with previous results for local ETGs 
 (Ko et al. 2009, 2012, 2013); most of more massive red ETGs with 
 $M_{star}$ $>$ 10$^{11.2}$ $M_{\odot}$ are in the quiescent mode, 
 but less massive red ETGs with $M_{star}$ $<$ 10$^{11.2}$ $M_{\odot}$ 
 show a wide spread in sSFR. 
If we take into account 24 $\mu$m undetected galaxies as well, the plot suggests that a
 significant fraction of them have mid-IR excess emission. 
Because of 24 $\mu$m flux density upper limit, it is not straightforward to compute exactly
 what fraction of red ETGs can be classified as `mid-IR excess' in the sample of 
 less massive ETGs. 
With the result for local red ETGs, these results suggest that `mid-IR excess' emission, 
 a good indicator of recent star formation, is not rare for red ETGs at $z <$ 1 
 (especially for less massive galaxies). 
This conclusion is consistent with the result in Ilbert et al. (2010) who found that 
 massive red ETGs are already in place at $z$ $\sim$ 1 while the intermediate-mass red ETGs
 are still forming at $z$ $<$ 1, leaving a trace of recent star formation.

\subsubsection{Fractions of Red Early-type Galaxies with Mid-IR Excess Emission}

In Figure 9, we plot the fraction of GOODS red ETGs with mid-IR excess emission 
 (defined in Fig. 8) in the sample of more massive galaxies with 
 $M_{star} > 10^{11.2} M_{\odot}$.
We use this mass cut so that all the 24 $\mu$m undetected galaxies have smaller sSFRs 
 than the sSFR cut for mid-IR excess. 
In other words, 24 $\mu$m undetected galaxies are quiescent systems without mid-IR excess
 emission.  
For comparison, we also show the fraction of SDSS red ETGs with mid-IR excess emission
 in the sample of massive galaxies using the same mass limit as for GOODS galaxies. 
We only use the galaxies at $0.04 < z < 0.06$ where the $WISE$ 22 $\mu$m detection limit 
 roughly corresponds to SFR $\sim$1 $M_{\odot}yr^{-1}$ (see Fig. 5). 
This limit then ensures that most $WISE$ 22 $\mu$m undetected galaxies in these mass and
 redshift ranges are quiescent systems without mid-IR excess emission 
 (i.e., log sSFR $\lesssim -11$, see Fig. 8), similar to GOODS galaxies. 
Despite the small redshift range, there are still some $WISE$ 22 $\mu$m undetected galaxies
 with SFR $>$ 1 $M_{\odot}yr^{-1}$ (e.g., see Fig. 5 for $z$ $\sim$ 0.05 galaxies). 
Therefore, the fraction for SDSS galaxies indicates a lower limit.

Figure 8 shows that the majority of massive red ETGs are in a `relatively' quiescent mode 
 at $z < 1$ even though some of them are detected in the mid-IR.
This is consistent with previous results that most massive quiescent galaxies (mainly 
 red ETGs) are created slowly at $z < 1$ (e.g., Borch et al. 2006; Bundy et al. 2006; 
 Arnouts et al. 2007; Ilbert et al. 2010, 2013; Cassata et al. 2011; Kajisawa et al. 2011;
 Moustakas et al. 2013).
For example, Ilbert et al. (2010) showed that more than 50\% of galaxies with 
 $M_{star} > 10^{11} M_{\odot}$ are quiescent at $z < 1$.
However, Figure 9 shows that 18-30\% of massive red ETGs can be classified
 as `mid-IR excess' galaxies even though many of massive red ETGs are in a `relatively' 
 quiescent mode.
This suggests that recent star formation is not negligible even in massive red ETGs at 
 $z < 1$; note that we conservatively consider that all massive red ETGs without mid-IR 
 detection at $z < 1$ have no excess emission.

\begin{figure}[!ht]
\centering
\includegraphics[width=15cm]{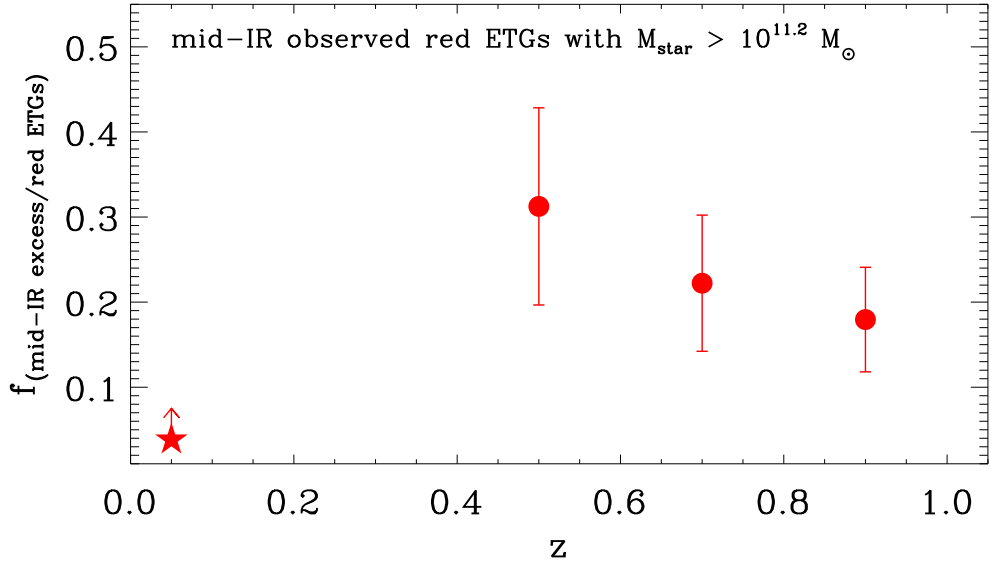}
\caption{Fraction of mid-IR excess galaxies among GOODS
         massive ($M_{star}$ $>$ 10$^{11.2}$ $M_{\odot}$) red ETGs (filled circles).
         Star symbols represent SDSS massive red ETGs.  Upward arrow indicates a lower 
         limit of the fraction of mid-IR excess galaxies. 
          }
         
\end{figure}

\section{CONCLUSIONS}

We study the mid-IR excess emission of early-type galaxies on the red sequence 
 at $z$ $<$ 1 using the spectroscopic sample of galaxies in the fields of GOODS. 
We find that most GOODS ETGs remain red in the $u-r$ colors, but a significant
 fraction of them show a wide range of NUV$-r$ and $r-$[12 $\mu$m] colors;
 this suggests that the fraction of young and intermediate-age stars in these massive 
 red ETGs with $M_{star}$ $>$ 10$^{10.5}$ $M_{\odot}$ is not negligible. 
The combination of near-UV and mid-IR provides not only an efficient tool to detect 
 ETGs with recent star formation, but also to break the degeneracy between 
 quiescent and dusty star-forming galaxies.

We also find that less massive ETGs are more likely to experience recent star formation
 or/and low-level ongoing star formation, consistent with the result of local red ETGs. 
This result is also consistent with the previous result that more massive galaxies have 
 lower sSFRs at $z$ $<$ 1 (Rodighiero et al. 2010; Karim et al. 2011). 
In other words, more massive galaxies (mainly quiescent, red ETGs) are already in place 
 at $z$ $\sim$ 1, but less massive galaxies continuously increase their stellar masses 
 through several physical processes such as minor mergers 
 (Kaviraj et al. 2009; Ilbert et al. 2010; Moustakas et al. 2013; Damjanov et al. 2014).

We explore the fraction of mid-IR excess galaxies among 
 massive ($M_{star} > 10^{11.2} M_{\odot}$) red ETGs at $z$ $<$ 1.  
We find that the majority of massive red ETGs are `quiescent' systems. 
However, more than 18\% of GOODS galaxies at 0.4 $<$ $z$ $<$ 1.0 still show 
 mid-IR excess emission even though we consider all the 24 $\mu$m undetected galaxies 
 to be quiescent systems without mid-IR excess emission.
These findings suggest that the recent star formation is not rare even among quiescent, 
 red ETGs at $z$ $<$ 1, if the mid-IR excess emission results mainly from 
 intermediate-age stars or/and low-level current star formation.

\begin{acknowledgements}

J.K. and J.C.L. are the members of Dedicated Researchers for Extragalactic AstronoMy (DREAM) 
 in Korea Astronomy and Space Science Institute (KASI).
H.S.H. acknowledges the Smithsonian Institution for the support of his post-doctoral
 fellowship.
M.I. acknowledges the support from the National Research Foundation of Korea (NRF) grant, 
 No. 2008-0060544, funded by the Korea government (MSIP).
D.L.B. acknowledges the French National Agency for Research (ANR) for their support 
(ANR-09-BLAN-0224).

\end{acknowledgements}

\bibliographystyle{apj} 

\begin{thebibliography}

\bibitem{} Abazajian, K. N., Adelman-McCarthy, J. K., Ag\"{u}eros, M. A., et al. 2009, \apjs,
182, 543
\bibitem{} Arnouts, S., Cristiani, S., Moscardini, L., et al.\ 1999, \mnras, 310, 540 
\bibitem{} Arnouts, S., Walcher, C.~J., Le F{\`e}vre, O., et al.\ 2007, \aap, 476, 137 
\bibitem{} Assef, R. J., Kochanek, C. S., Brodwin, M., et al. 2010, \apj, 713, 970 
\bibitem{} Baldry, I. K., Glazebrook, K., Brinkmann, J., et al. 2004, \apj, 600, 681
\bibitem[Balestra et al.(2010)]{bal10} 
  Balestra, I., Mainieri, V., Popesso, P., et al.\ 2010, \aap, 512, A12 
\bibitem{} Balogh, M., Eke, V., Miller, C., et al. 2004, \mnras, 348, 1355
\bibitem{} Bamford, S. P., Nichol, R. C.; Baldry, I. K., et al. 2009, \mnras, 393, 1324
\bibitem[Barger et al.(2008)]{bar08} 
  Barger, A.~J., Cowie, L.~L., \& Wang, W.-H.\ 2008, \apj, 689, 687 
\bibitem{} Bell, E. F., Wolf, C., Meisenheimer, K., et al. 2004, \apj, 608, 752
\bibitem{} Blanton, M. R., Hogg, D. W., Bahcall, N. A., et al. 2003, \apj, 594, 186
\bibitem{} Borch, A., Meisenheimer, K., Bell, E.~F., et al.\ 2006, \aap, 453, 869 
\bibitem{} Bressan, A., Granato, G. L., \& Silva, L. 1998, \aap, 332, 135
\bibitem{} Bressan, A., Panuzzo, P., Buson, L., et al. 2006, \apj, 639, L55
\bibitem{} Bruzual, G., \& Charlot, S. 2003, \mnras, 344, 1000
\bibitem{} Buitrago, F., Trujillo, I., Conselice C. J., et al. 2013, \mnras, 428, 1460
\bibitem[Bundy et al.(2005)]{bun05} 
  Bundy, K., Ellis, R.~S., \& Conselice, C.~J.\ 2005, \apj, 625, 621 
\bibitem{} Bundy, K., Ellis, R.~S., Conselice, C.~J., et al.\ 2006, \apj, 651, 120 
\bibitem{} Calzetti, D., Armus, L., Bohlin, R.~C., et al.\ 2000, \apj, 533, 682 
\bibitem{} Cassata, P., Giavalisco, M., Guo, Y., et al.\ 2011, \apj, 743, 96 
\bibitem[Cervantes-Sodi et al.(2012)]{2012MNRAS.426.1606C} Cervantes-Sodi, B., Hernandez, X.,
  Hwang, H.~S., Park, C., \& Le Borgne, D.\ 2012, \mnras, 426, 1606 
\bibitem[Choi et al.(2009)]{2009MNRAS.395..637C} Choi, Y., Goto, T., 
\& Yoon, S.-J.\ 2009, \mnras, 395, 637 
\bibitem{} Choi, Y.-Y., Park, C., \& Vogeley, M. S. 2007, \apj, 658, 884
\bibitem{} Choi, Y.-Y., Han, D., \& Kim, S. S. 2010, J. Korean Astron. Soc., 43, 191
\bibitem{} Chary, R., \& Elbaz, D. 2001, \apj, 556, 562
\bibitem{} Cimatti, A., Daddi, E., \& Renzini, A.\ 2006, \aap, 453, L29 
\bibitem{} Clemens, M. S., Bressan, A., Panuzzo, P., et al. 2009, \mnras, 392, 982
\bibitem[Cohen et al.(2000)]{coh00} 
  Cohen, J.~G., Hogg, D.~W., Blandford, R., et al.\ 2000, \apj, 538, 29 
\bibitem[Cooper et al.(2011)]{coo11} 
  Cooper, M.~C., Aird, J.~A., Coil, A.~L., et al.\ 2011, \apjs, 193, 14 
\bibitem[Cooper et al.(2012)]{coo12} 
  Cooper, M.~C., Yan, R., Dickinson, M., et al.\ 2012, \mnras, 425, 2116 
\bibitem{} Cowie, L.~L., Songaila, A., Hu, E.~M., \& Cohen, J.~G.\ 1996, \aj, 112, 839
\bibitem[Cowie et al.(2004)]{cow04} 
  Cowie, L.~L., Barger, A.~J., Hu, E.~M., Capak, P., \& Songaila, A.\ 2004, \aj, 127, 3137 
\bibitem[Damjanov et al.(2014)]{2014arXiv1405.2934D} Damjanov, I., Hwang, 
H.~S., Geller, M.~J., \& Chilingarian, I.\ 2014, arXiv:1405.2934 
\bibitem{} Dekel, A., Birnboim, Y., Engel, G., et al. 2009, \nat, 457, 451
\bibitem{} Dickinson, M., Giavalisco, M., \& GOODS Team 2003, The Mass of Galaxies at Low and High Redshift, 324 
\bibitem{} Donoso, E., Yan, L., Tsai, C., et al. 2012, \apj, 748, 80
\bibitem{} Elbaz, D., Hwang, H.~S., Magnelli, B., et al.\ 2010, \aap, 518, L29 
\bibitem[{Elbaz et al.(2011){Elbaz}, {Dickinson}, {Hwang},
  {D{\'{\i}}az-Santos}, {Magdis}, {Magnelli}, {Le Borgne}, {Galliano},
  {Pannella}, {Chanial}, {Armus}, {Charmandaris}, {Daddi}, {Aussel}, {Popesso},
  {Kartaltepe}, {Altieri}, {Valtchanov}, {Coia}, {Dannerbauer}, {Dasyra},
  {Leiton}, {Mazzarella}, {Alexander}, {Buat}, {Burgarella}, {Chary}, {Gilli},
  {Ivison}, {Juneau}, {Le Floc'h}, {Lutz}, {Morrison}, {Mullaney}, {Murphy},
  {Pope}, {Scott}, {Brodwin}, {Calzetti}, {Cesarsky}, {Charlot}, {Dole},
  {Eisenhardt}, {Ferguson}, {F{\"o}rster Schreiber}, {Frayer}, {Giavalisco},
  {Huynh}, {Koekemoer}, {Papovich}, {Reddy}, {Surace}, {Teplitz}, {Yun}, \&
  {Wilson}}]{elb11} 
{Elbaz}, D., {Dickinson}, M., {Hwang}, H.~S., {et~al.} 2011, \aap, 533, 119
\bibitem{} Faber, S. M., Willmer, C. N. A., Wolf, C., et al. 2007, \apj, 665, 265
\bibitem{} Ferreras I., \& Silk J. 2000, \apj, 541, L37
\bibitem[{{Fioc} \& {Rocca-Volmerange}(1999)}]{fr99}
{Fioc}, M. \& {Rocca-Volmerange}, B. 1999, arXiv:astro-ph/9912179
\bibitem{} Fritz, A., Scodeggio, M., Ilbert, O., et al. 2014, \aap, 563, A92 
\bibitem{} Fukugita, M., Nakamura, O., Turner, E.~L., Helmboldt, J., \& Nichol, R.~C.
\ 2004, \apjl, 601, L127 
\bibitem{} Gallazzi, A., Charlot, S., Brinchmann, J., White, S.~D.~M., 
\& Tremonti, C.~A. 2005, \mnras, 362, 41 
\bibitem{} Gallazzi, A., Bell, E. F., Wolf, C., et al. 2009, \apj, 690, 1883
\bibitem{} Giavalisco, M., Ferguson, H.~C., Koekemoer, A.~M., et al.\ 2004, \apjl, 600, L93 
\bibitem[Grogin et al.(2011)]{gro11} 
  Grogin, N.~A., Kocevski, D.~D., Faber, S.~M., et al.\ 2011, \apjs, 197, 35 
\bibitem{} Hopkins, P.~F., Hernquist, L., Cox, T.~J., et al.\ 2006, \apjs, 163, 1 
\bibitem{} Hopkins, P.~F., Bundy, K., Croton, D., et al.\ 2010, \apj, 715, 202 
\bibitem{} Huertas-Company, M., Mei, S., Shankar, F., et al.\ 2013, \mnras, 428, 1715 
\bibitem[{{Hwang} \& {Park}(2009)}]{hp09}
  {Hwang}, H.~S. \& {Park}, C. 2009, \apj, 700, 791
\bibitem{} Hwang, H.~S., Elbaz, D., Magdis, G., et al. 2010a, \mnras, 409, 75 
\bibitem{} Hwang, H. S., Elbaz, D., Lee, J. C., et al. 2010b, \aap, 522, A33
\bibitem[Hwang et al.(2011)]{hwa11goods} 
  Hwang, H.~S., Elbaz, D., Dickinson, M., et al.\ 2011, \aap, 535, A60 
\bibitem{} Hwang, H. S., Geller, M. J., Diaferio, A., \& Rines, K. J. 2012a, \apj, 752, 64
\bibitem[Hwang et al.(2012)]{2012ApJ...758...25H} Hwang, H.~S., Geller, M.~J., Kurtz, M.~J., 
  Dell'Antonio, I.~P., \& Fabricant, D.~G.\ 2012b, \apj, 758, 25 
\bibitem{} Ilbert, O., Arnouts, S., McCracken, H.~J., et al.\ 2006, \aap, 457, 841 
\bibitem{} Ilbert, O., Salvato, M., Le Floc'h, E., et al. 2010, \apj, 709, 644
\bibitem{} Ilbert, O., McCracken, H.~J., Le F{\`e}vre, O., et al.\ 2013, \aap, 556, A55 
\bibitem{} Im, M., Simard, L., Faber, S. M., et al. 2002, \apj, 571, 136
\bibitem{} Jarrett, T. H., Cohen, M., Masci, F., et al. 2011, \apj, 735, 112
\bibitem{} Kajisawa, M., Ichikawa, T., Yoshikawa, T., et al. 2011, \pasj, 63, 403
\bibitem{} Karim, A., Schinnerer, E., Mart{\'{\i}}nez-Sansigre, A., et al. 2011, \apj, 730, 61
\bibitem{} Kauffmann, G., Heckman, T. M., White, S. D. M., et al. 2003, \mnras, 341, 54
\bibitem[Kaviraj et al.(2007)]{2007MNRAS.382..960K} Kaviraj, S., Kirkby, 
L.~A., Silk, J., \& Sarzi, M.\ 2007a, \mnras, 382, 960 
\bibitem{} Kaviraj, S., Schawinski, K., Devriendt, J. E. G., et al. 2007b, \apjs, 173, 619
\bibitem{} Kaviraj, S., Peirani, S., Khochfar, S., Silk, J., \& Kay, S.\ 2009, \mnras, 394,
 1713 
\bibitem{} Kelson, D. D., \& Holden, B. P. 2010, \apj, 713, L28
\bibitem{} Kennicutt, Jr., R. C. 1998, ARA\&A, 36, 189
\bibitem{} Khochfar, S., \& Silk, J. 2006, \apj, 648, L21
\bibitem{} Ko, J., Im, M., Lee, H. M., et al. 2009, \apj, 695, L198
\bibitem{} Ko, J., Im, M., Lee, H. M., et al. 2012, \apj, 745, 181
\bibitem{} Ko, J., Hwang, H. S., Lee, J. C., \& Sohn, Y.-J. 2013, \apj, 767, 90
\bibitem[Koekemoer et al.(2011)]{koe11} Koekemoer, A.~M., 
  Faber, S.~M., Ferguson, H.~C., et al.\ 2011, \apjs, 197, 36 
\bibitem{} Kodama, T., Bower, R.~G., \& Bell, E.~F.\ 1999, \mnras, 306, 561 
\bibitem[Kurk et al.(2009)]{kurk09} 
  Kurk, J., Cimatti, A., Zamorani, G., et al.\ 2009, \aap, 504, 331 
\bibitem{} Lacy, M., Storrie-Lombardi, L.~J., Sajina, A., et al.\ 2004, \apjs, 154, 166 
\bibitem[Le Borgne \& Rocca-Volmerange(2002)]{lr02}
  {Le Borgne}, D. \& {Rocca-Volmerange}, B. 2002, \aap, 386, 446
\bibitem[Le F{\`e}vre et al.(2004)]{lef04} 
  Le F{\`e}vre, O., Vettolani, G., Paltani, S., et al.\ 2004, \aap, 428, 1043 
\bibitem{} Lee, J.~H., Lee, M.~G., \& Hwang, H.~S.\ 2006, \apj, 650, 148
\bibitem{} Lee, J.~H., Hwang, H.~S., Lee, M.~G., Lee, J.~C., \& Matsuhara, H.
\ 2010, \apj, 719, 1946 
\bibitem{} Lee, J. C., Hwang, H. S., Lee, M. G., et al. 2012, \apj, 756, 95
\bibitem{} Lee, J.~C., Hwang, H.~S., \& Ko, J.\ 2013, \apj, 774, 62 
\bibitem{} Magnelli, B., Elbaz, D., Chary, R. R., et al. 2009, \aap, 496, 57
\bibitem[Magnelli et al.(2011)]{mag11} 
  Magnelli, B., Elbaz, D., Chary, R.~R., et al.\ 2011, \aap, 528, A35 
\bibitem{} Martin, D. C., Fanson, J., Schiminovich, D., et al. 2005, \apj, 619, L1
\bibitem{} Martini, P., Dicken, D., \& Storchi-Bergmann, T.\ 2013, \apj, 766, 121
\bibitem{} Masters, K. L., Mosleh, M., Romer, A. K., et al. 2010, \mnras, 405, 783
\bibitem[Mignoli et al.(2005)]{mig05} 
  Mignoli, M., Cimatti, A., Zamorani, G., et al.\ 2005, \aap, 437, 883 
\bibitem{} Moustakas, J., Coil, A.~L., Aird, J., et al.\ 2013, \apj, 767, 50 
\bibitem{} Naab, T., Johansson, P.~H., \& Ostriker, J.~P.\ 2009, \apjl, 699, L178 
\bibitem{} Newman, A.~B., Ellis, R.~S., Bundy, K., \& Treu, T.\ 2012, \apj, 746, 162 
\bibitem{} Piovan, L., Tantalo, R., \& Chiosi, C. 2003, \aap, 408, 559
\bibitem{} Polletta, M., Tajer, M., Maraschi, L., et al. 2007, \apj, 663, 81
\bibitem[Popesso et al.(2009)]{pop09} 
  Popesso, P., Dickinson, M., Nonino, M., et al.\ 2009, \aap, 494, 443 
\bibitem[Ravikumar et al.(2007)]{rav07} 
  Ravikumar, C.~D., Puech, M., Flores, H., et al.\ 2007, \aap, 465, 1099 
\bibitem[Reddy et al.(2006)]{red06} 
  Reddy, N.~A., Steidel, C.~C., Erb, D.~K., Shapley, A.~E., \& Pettini, M.\ 2006, \apj, 653, 1004 
\bibitem{} Rodighiero, G., Cimatti, A., Gruppioni, C., et al. 2010, \aap, 518, L25
\bibitem{} Salim, S., Dickinson, M., Michael R. R., et al. 2009, \apj, 700, 161 
\bibitem{} Salim, S., \& Rich, M. 2010, \apj, 714, L290
\bibitem{} Salim, S., Fang, J.~J., Rich, R.~M., Faber, S.~M., \& Thilker, D.~A.\ 2012, \apj, 755, 105
\bibitem[Salpeter(1955)]{sal55} 
  Salpeter, E.~E. 1955, \apj, 121, 161
\bibitem{} Schawinski, K., Kaviraj, S., Khochfar, S., et al. 2007, \apjs, 173, 512
\bibitem{} Shim, H., Im, M., Lee, H. M., et al. 2011, \apj, 727, 14
\bibitem{} Silva, L., Granato, G. L., Bressan, A., \& Danese, L. 1998, \apj, 509, 103
\bibitem[Silverman et al.(2010)]{sil10} 
  Silverman, J.~D., Mainieri, V., Salvato, M., et al.\ 2010, \apjs, 191, 124 
\bibitem{} Skibba, R.~A., Engelbracht, C.~W., Dale, D., et al.\ 2011, \apj, 738, 89 
\bibitem{} Stern, D., Eisenhardt, P., Gorjian, V., et al.\ 2005, \apj, 631, 163 
\bibitem{} Strateva, I., Ivezi\'{c}, \v{Z}, \& Knapp, G. R. 2001, \aj, 122, 1861
\bibitem[Szokoly et al.(2004)]{szo04} 
  Szokoly, G.~P., Bergeron, J., Hasinger, G., et al.\ 2004, \apjs, 155, 271 
\bibitem[Teplitz et al.(2011)]{tep11} 
  Teplitz, H.~I., Chary, R., Elbaz, D., et al.\ 2011, \aj, 141, 1 
\bibitem{} Tinker, J. L., Leauthaud, A., Bundy, K., et al. 2013, \apj, 778, 93
\bibitem[Vanzella et al.(2005)]{van05} 
  Vanzella, E., Cristiani, S., Dickinson, M., et al.\ 2005, \aap, 434, 53 
\bibitem[Vanzella et al.(2006)]{van06} 
  Vanzella, E., Cristiani, S., Dickinson, M., et al.\ 2006, \aap, 454, 423 
\bibitem[Vanzella et al.(2008)]{van08} 
  Vanzella, E., Cristiani, S., Dickinson, M., et al.\ 2008, \aap, 478, 83 
\bibitem{} Vega, O., Bressan, A., Panuzzo, P., et al. 2010, \apj, 721, 1090
\bibitem{} Willmer, C. N. A., Faber, S. M., Koo, D. C., et al. 2006, \apj, 647, 853
\bibitem[Wirth et al.(2004)]{wir04} 
  Wirth, G.~D., Willmer, C.~N.~A., Amico, P., et al.\ 2004, \aj, 127, 3121 
\bibitem{} Wolf, C., Gray, M.~E., \& Meisenheimer, K.\ 2005, \aap, 443, 435 
\bibitem{} Wolf, C., Arag\'{o}n-Salamanca, A., Balogh, M., et al. 2009, \mnras, 393, 1302
\bibitem[Xia et al.(2011)]{xia11} 
  Xia, L., Malhotra, S., Rhoads, J., et al.\ 2011, \aj, 141, 64 
\bibitem{} Yi, S. K., Yoon, S.-J., Kaviraj, S., et al. 2005, \apj, 619, L111
\bibitem{} York, D. G., Adelman, J., Anderson, J. E. Jr., et al. 2000, \aj, 120, 1579

\end{thebibliography}

\clearpage

\end{document}